\documentclass[11pt]{article}

\usepackage{xspace}                                     %
\usepackage{microtype}  %

\usepackage{fullpage}
\usepackage{amsfonts, amsthm, mathtools,amssymb}
\usepackage{algorithm}
\usepackage[noend]{algpseudocode}
\usepackage[colorlinks,citecolor=blue,bookmarks=true,linktocpage]{hyperref}
\usepackage{aliascnt}
\usepackage[numbered]{bookmark}
\usepackage[capitalise]{cleveref}

\usepackage[T1]{fontenc}
\usepackage{palatino}
\makeatletter
 \@ifundefined{theorem}{%
   \theoremstyle{plain} %
   	\newtheorem{theorem}{Theorem}[section]
   	\newtheorem*{theorem*}{Theorem} %
   	\newaliascnt{coro}{theorem}
   	  
   	\aliascntresetthe{coro}
   	\newaliascnt{lem}{theorem}
   		\newtheorem{lemma}[lem]{Lemma}
   	\aliascntresetthe{lem}
   	\newaliascnt{clm}{theorem}
   		
 	\aliascntresetthe{clm}
 	\newaliascnt{fact}{theorem}

 	\aliascntresetthe{fact}
   	
   \newaliascnt{prop}{theorem}
   		
 	\aliascntresetthe{prop}
 	\newaliascnt{conj}{theorem}
   		
 	\aliascntresetthe{conj}

   \theoremstyle{remark} %

   \theoremstyle{definition} %
   	\newaliascnt{defn}{theorem}
  		 \newtheorem{definition}[defn]{Definition}
  	 \aliascntresetthe{defn}

}{}
\makeatother

\providecommand{\email}[1]{\href{mailto:#1}{\nolinkurl{#1}\xspace}}

\newcommand{\rQuantile}{\mathtt{rQuantile}}
\newcommand{\rMedian}{\mathtt{rMedian}}

\newcommand{\lcakp}{\mathtt{LCA-KP}}
\newcommand{\mappinggreedy}{\mathtt{MAPPING-GREEDY}}
\newcommand{\greedyKP}{\mathtt{GREEDY}}
\newcommand{\efficiencythreshold}{\mathtt{CONVERT-GREEDY}}

\newcommand{\eps}{\varepsilon}

\usepackage{tikz}
\usepackage[backref=true,backend=biber,style=alphabetic,maxbibnames=99,maxcitenames=99,maxalphanames=99]{biblatex}
\newcommand{\OPT}{\ensuremath{\operatorname{OPT}}\xspace}

\begin{document}

\title{Local Computation Algorithms for Knapsack: impossibility results, and how to avoid them}

\author{
Cl\'ement L. Canonne\thanks{The University of Sydney.
Email: \email{clement.canonne@sydney.edu.au}.}
\and
Yun Li\thanks{The University of Sydney.
Email: \email{yunli97@outlook.com}.}
\and 
Seeun William Umboh\thanks{The University of Melbourne, ARC Training Centre in Optimisation Technologies, Integrated Methodologies, and Applications (OPTIMA). Email: \email{william.umboh@unimelb.edu.au}.}
}

\maketitle

\begin{abstract}
Local Computation Algorithms (LCA), as introduced by Rubinfeld, Tamir, Vardi, and Xie~(2011), are a type of ultra-efficient algorithms which, given access to a (large) input for a given computational task, are required to provide fast query access to a consistent output solution, without maintaining a state between queries. This paradigm of computation in particular allows for hugely distributed algorithms, where independent instances of a given LCA provide consistent access to a common output solution.

The past decade has seen a significant amount of work on LCAs, by and large focusing on graph problems. In this paper, we initiate the study of Local Computation Algorithms for perhaps \emph{the} archetypal combinatorial optimization problem, Knapsack. We first establish strong impossibility results, ruling out the existence of any non-trivial LCA for Knapsack as several of its relaxations. We then show how equipping the LCA with additional access to the Knapsack instance, namely, weighted item sampling, allows one to circumvent these impossibility results, and obtain sublinear-time and query LCAs. Our positive result draws on a connection to the recent notion of \emph{reproducibility} for learning algorithms (Impagliazzo, Lei, Pitassi, and Sorrell,~2022), a connection we believe to be of independent interest for the design of LCAs.
\end{abstract}

\section{Introduction}
Most classical algorithms can be divided into three phases: (1)~read and pre-process the entire input; (2)~perform the computations to solve the problem; and (3)~finally output the solution. However, several constraints can impact the effectiveness of these algorithms in practice. For instance, computation might become infeasible with massive size input that could not fit in memory; finding an optimal solution may be exceptionally difficult while a quick response is needed; or the size of the solution itself might be too large to allow for it to be easily outputted. Different algorithmic paradigms have been introduced to address these issues, e.g., the streaming model~\cite{AlonMS96} to address memory constraints, or property testing~\cite{GoldreichGR96} to address very large inputs in the context of sublinear-time algorithms.

Local Computation Algorithms (LCAs) are sublinear time and space algorithms for search problems first introduced by~\cite{rubinfeld2011fast}. The LCA model seeks to capture settings where \emph{both} input and output are massive, so even \emph{efficient} algorithms are too time-consuming, as both the time required to fully read the input and describe the output are too large. LCAs implement efficient query access to a small portion of a valid solution of the underlying problem without computing the whole output, while maintaining no state between the various queries it is asked to answer (for a full definition of the model, see~\cref{sec:preliminaries}). Due to this last requirement, LCAs are particularly well-suited for the distributed and parallel settings, as they allow for many instances of the algorithm to be run independently, each providing local query access to the same solution to the computational problem at hand. 

Most of the work on LCAs, since their introduction, has focused on graph problems, such as Hypergraph Coloring, Independent Set Cover, Maximal Independent Set~\cite{rubinfeld2011fast,AlonRVX12,Ghaffari22}, and Maximum Matching~\cite{LeviRY17,BehnezhadRR23}, to name a few. Yet, far fewer works have considered other computational tasks in the context of LCAs, and in particular combinatorial optimization questions. In this work, we revisit this state of affairs, focusing on one of the most fundamental combinatorial optimization problems: Knapsack. In this setting, the algorithm is provided with a (read-only) random seed, and given query access to an instance $I$ of Knapsack: upon receiving query asking whether item $i$ is part of an optimal solution, the algorithm is allowed to make a small (sublinear in the instance size) number of queries to $I$, and must answer according to some feasible solution $S=S(I)$ to Knapsack (with high probability). The algorithm must be able to answer as many such queries as desired, in arbitrary order and without maintaining a state between queries, while providing consistent access to the same solution $S$.

\subsection{Our results and contributions}
    \label{sec:intro:results}
Our first set of results investigates the very possibility of obtaining an efficient local computation algorithm for Knapsack. Our first theorem rules out any sublinear-time\footnote{In what follows, and in particular our lower bounds, we ignore the computational complexity aspects of the algorithms and focus on their \emph{query complexity}, that is, the worst-case number of queries to the input it needs to perform in order to answer an LCA query about the output. The query complexity clearly lower bounds time complexity; but this distinction is important to avoid vacuous conditional results (assuming $\textsf{P}\neq\textsf{NP}$), as the search version of Knapsack is NP-Hard.} LCA who provides query access to an optimal solution:

\begin{theorem}[Informal; see~\cref{theo:no:optimal:knapsack:lca}]
    \label{theo:no:optimal:knapsack:lca:informal}
    There is no sublinear-time LCA for Knapsack that provides consistent query access to an optimal solution.
\end{theorem}
This hardness result, however, only rules out algorithms for the \emph{exact} version of Knapsack: one could still hope for non-trivial approximation algorithms in the local computation model. Our second result closes this door as well:
\begin{theorem}[Informal; see~\cref{theo:no:approx:knapsack:lca}]
    \label{theo:no:approx:knapsack:lca:informal}
    There is no sublinear-time LCA for Knapsack that provides consistent query access to an $\alpha$-approximate solution, for any fixed $\alpha\in(0,1]$.
\end{theorem}
One could still relax the goal in another direction: instead of requiring optimality or near-optimality of the solution provided by the LCAs, one could ask for access to any \emph{maximal} feasible solution, regardless of its value~--~that is, any solution which cannot be improved by adding further items. %
Unfortunately, we show that even this somewhat modest goal is impossible:
\begin{theorem}[Informal; see~\cref{theo:no:maximal:knapsack:lca}]
    \label{theo:no:maximal:knapsack:lca:informal}
    There is no sublinear-time LCA for Knapsack that provides consistent query access to a maximal feasible solution.
\end{theorem}
These strong impossibility results, at first glance, seem to close our line of inquiry: any LCA providing access to a reasonable solution of Knapsack must essentially query the whole input. But this relies on the fact that the algorithm has only very limited access to the instance of the problem: also random query access to the instance $I$ might seem powerful, this also does not allow it to easily leverage any feature of the instance, as the weights and profits across items are essentially independent. One avenue to circumvent these impossibility results is to equip the algorithms with a stronger and natural type of query access to $I$: following~\cite{ito2012constant} (in the classical setting), we consider a \emph{weighted sampling} model, where the LCA can randomly sample items from the instance $I$ \emph{proportionally to their profit}. Intuitively, this should enable the algorithm to focus on the most relevant items, which are more likely to be sampled. We show that this is indeed the case, and are able to obtain a query-efficient LCA for Knapsack in this weighted sampling model:
\begin{theorem}[Main theorem (Informal; see~\cref{theo:lcakpresult})]\label{theo:lcakpresult:informal}
    There exists an LCA which, given weighted sampling access to the Knapsack instance, provides consistent query access to a $(1/2+\eps)$-approximation, for any fixed $\eps > 0$. The LCA has query complexity
    \[
    \left( 1/\eps \right)^{O(\log^{*}n)}\;,
    \]
    where $\log^\ast$ denotes the iterated logarithm.
\end{theorem}
\noindent We elaborate on the techniques, and provide an outline of the proofs, below.

\paragraph{Technical overview.} Our first two impossibility results,~\cref{theo:no:optimal:knapsack:lca:informal,theo:no:approx:knapsack:lca:informal}, follow a very similar approach: namely, we reduce the Knapsack problem on $n$ items to computing the OR function on $n$ bits, before invoking the known randomized query complexity lower bound for $\textsf{OR}_n$. While the resulting reduction itself is quite simple, the key aspect is to maintain the type of query access provided (i.e., simulate query access to the constructed Knapsack instance $I$ given query access to the input $x\in\{0,1\}^n$ for the $\textsf{OR}_n$ function) in a local way: that is, for the lower bound to carry through with no loss in parameters, we want to be able to answer any LCA query to $I$ with only a small number~--~ideally $O(1)$~--~of queries to $x$.

The third impossibility result,~\cref{theo:no:maximal:knapsack:lca:informal}, is more involved, and does not follow from a reduction. Instead, we prove it directly, first restricting ourselves to deterministic algorithms using Yao's Principle, and defining a suitable distribution of ``hard Knapsack instances.'' These hard instances have the following structure: the weight limit is $K=1$, and all items, except for a uniformly randomly chosen pair, have weight zero (recall that since we consider the maximal feasibility version of Knapsack in this theorem, we do not have to define or care about the items' profits). The remaining two, items $i$ and $j$, have weights randomly chosen to be either $3/4, 1/4$ or $3/4, 3/4$. Note that in the first case, a maximally feasible solution must include both items $i,j$; while in the second case, only one of the two can~--~and must~--~be part of the solution. The crux of the argument is then to use both the consistency and memorylessness requirements of an LCA to show that, unless it makes enough queries to the input to find the (randomly chosen) location of both special items $i,j$, any LCA must err on some sequence of (constantly many) queries, either by including both items when it should include only one, or only one when it should include both. This gives an $\Omega(n)$ lower bound for the number of queries required to answer such an $O(1)$-length sequence of queries, and so an $\Omega(n)$ lower bound on the query complexity.

The starting point for our positive result,~\cref{theo:lcakpresult:informal}, is the work of Ito, Kiyoshima, and Yoshida~\cite{ito2012constant}, which provides a constant-time randomized algorithm for approximating the optimal \emph{value} of a solution to a Knapsack instance. At a high level, given a parameter $\eps$ their algorithm works by first (implicitly) partitioning the $n$ items, each given by a profit $p_i$ and weight $w_i$, in 3 sets: (1) the set $L$ of large items, with profit $p_i> \eps^2$; (2) the set $S$ of small items, with profit $p_i \leq \eps^2$ but large ``efficiency ratio'' $p_i/w_i \geq \eps^2$; and (3)~the set $G$ of garbage items, with low profit and low efficiency. (To interpret the notion of efficiency, recall the greedy algorithm for Fractional Knapsack, which first sorts items by non-increasing efficiency ratio before selecting as many as possible in this order.) The algorithm of~\cite{ito2012constant} then samples $\tilde{O}(1/\eps^4)$ items proportionally to their profit, discarding all items from $G$: by a coupon collector argument, with high probability all items from $L$ will be sampled, and further the algorithm will obtain a good approximation of the frequency distribution of items in $S$: a set of efficiency thresholds $e_1,\dots, e_k$, for $k=O(1/\eps)$, with the guarantee that for each $\ell$ there is approximately an $\eps$ fraction of the total profit on items with efficiency ratio $\approx e_\ell$. Using these two facts, the algorithm can build a new \emph{constant-size} (i.e., $O_\eps(1)$-size) instance $I'$ to Knapsack by including all large items and a suitable number of ``representatives'' for each efficiency from the small items, whose optimal value $\eps$-approximates that of the original instance, and solve this new instance optimally (in exponential time in the size of the new instance) before returning its value.

To adapt this approach to the LCA setting, we must overcome several obstacles. The first is that the above algorithm does not actually solve the original instance of Knapsack (which would be a hard problem), nor an approximation of it: it solves a different, constructed instance, which just happens to have approximately the same optimal value. But our LCA needs to answer very specific queries on the \emph{original} instance: \emph{``is item $i$ in the approximately optimal solution?''} While running the algorithm of  Ito, Kiyoshima, and Yoshida to obtain an optimal solution to $I'$ would let us answer this type of queries for ``large'' items (as these are included verbatim in the constructed instance $I'$)  and ``garbage'' items (as these are discarded, and not part of any solution), it does not readily let us get this information for any of the ``small'' items. To address this, we leverage the structure of the standard $1/2$-approximation algorithm for Knapsack (which takes the best of two solutions: that returned by the greedy algorithm mentioned above, and the singleton solution obtained by including the first item left out by this greedy algorithm). By running this $1/2$-approximation algorithm on the constructed instance $I'$, we end up with a threshold $\tau$ and very simple rule to decide whether a (constructed) item $i'$ is in the approximate solution to the (constructed) solution $I'$: check its efficiency ratio against this threshold $\tau$. But this threshold, by construction, also allows us (after carefully handling some corner cases) to decide whether an item $i$ of the \emph{original} instance is in a good feasible solution of the \emph{original} instance $I$: check if it is a large item (then we can decide easily if it is in the solution), a garbage item (then it is not), and otherwise it is a small item: check its efficiency ratio against $\tau$.

The above outline almost gets us where we want, and \emph{would} lead (if correct) to a $\operatorname{poly}(1/\eps)$-query LCA for $(2+\eps)$-approximate Knapsack. Unfortunately, it has one major issue, leading to our second obstacle: the sampling step used to approximate the distribution of efficiency profiles of the small items needs to be performed for each query, as the LCA is not allowed to maintain state between queries,  \emph{this random sampling will lead to inconsistent answers}. Indeed, even small variations in the efficiency thresholds $e_1,\dots, e_k$ computed from this sampling step may lead to a different threshold $\tau$, and as a result to our LCA failing to answer consistently to a single feasible solution. The solution would be to somehow have our sampling step give the same results across repetitions, which is of course not possible: one can easily design examples where randomly sampling items gives different outcomes with high probability, even across only a few repetitions; and naive attempts at rounding the sampled values suffer the same issue.

To solve this second major hurdle, we take recourse in the recent framework of \emph{reproducible learning} (Impagliazzo, Lei, Pitassi, and Sorrell~\cite{impagliazzo2022reproducibility}), originally proposed to ensure that learning algorithms output (with high probability) the same answer when run on a fresh sample of training data. Making this connection between the LCA consistency requirement and the reproducible learning definition, we are able to leverage the \emph{reproducible median} algorithm of~\cite{impagliazzo2022reproducibility} (suitably generalized to quantiles instead of simply median) to perform the frequency estimation step of our algorithm in a consistent fashion. Combining these building blocks together then yields our final algorithm: the $\log^\ast n$ dependence stemming from the use of the reproducible median, which provably requires this dependence on the domain size.

\subsection{Related Work}

\paragraph{Local Computation Algorithms.}
Since the introduction of Local Computation Algorithms in~\cite{rubinfeld2011fast}, many classical graph problems have been studied under this lens~\cite{rubinfeld2011fast,AlonRVX12,Ghaffari22,LeviRY17,grunau2020improved,BehnezhadRR23}, both from the upper and lower bound sides.

There is a strong connection between distribution algorithms and LCAs, which follows from earlier work by Parnas and Ron~\cite{parnas2007approximating}: broadly speaking, any $d$-round LOCAL algorithm with degree bound $\Delta$ can be simulated by a $q$-query LCA, where $q=O(\Delta^d)$. 

Another important design technique for LCAs (and one our algorithms leverage) is that of simulating greedy algorithms. For instance, \cite{nguyen2008constant} showed that some subclass of greedy algorithms for graph problems can be simulated by first randomly assigning a random number to the vertices, inducing a random ordering to the vertices. This technique was later generalized and improved by \cite{yoshida2012improved}. The connection to LCAs was first made explicit in~\cite{mansour2012converting}, which used it to show how to convert some online algorithm to LCAs while preserving the same approximation ratio. This led, among others, to LCAs for maximal matching with $O(\log^3n)$ time and space complexity and several load balancing problems with $O(\log n)$ time and space complexity. 

We conclude this short review of LCAs by mentioning the recent work of Biswas, Cao, Pyne, and Rubinfeld~\cite{BiswasCPR24}, which introduced a variant of the LCA model where the input is assumed to come from some probabilistic process (e.g., random Erd\H{o}s--R\'enyi graph). It would be interesting to see if such average-case assumptions could lead to faster LCAs for Knapsack on ``almost all instances,'' or even allow one to bypass our impossibility results.

\paragraph{Knapsack.}
The Knapsack problem is perhaps the archetypal combinatorial optimization problem, and one of Karp's original 21 NP-Hard problems. Many variants and relaxations have been studied over the years, including under stochastic assumptions, or various restrictions on the input. Given the vast body of work on Knapsack, we only discuss below the two most relevant to this paper, and refer the interested reader to~\cite[Section~3.1]{williamson2011design} for a more extensive coverage of the question and some of its approximation algorithms.

One important relaxation is when the quantity of each item to be included, instead of being binary, is allowed to take any value in $[0,1]$. This is known as the \emph{Fractional Knapsack}, and can be solved optimally by either linear programming or a very simple greedy algorithm which first sorts items by non-increasing ratio of $p_i/w_i$ (profit-to-weight), and then greedily picks items in this order until the Knapsack capacity is exhausted. This greedy algorithm, while in itself does not yield any non-trivial approximation ratio for the original Knapsack problem, can be easily modified to yield a $1/2$-approximation: see, e.g.,~\cite[Exercise~3.1]{williamson2011design}. We will draw on this simple $1/2$-approximation algorithm for our main algorithmic result. The second result on Knapsack we will rely on is the aforementioned work of Ito, Kiyoshima, and Yoshida~\cite{ito2012constant}, which, for any constant $\eps >0$, shows how to approximate (with high probability) the value of an optimal solution to a Knapsack instance to an additive $\pm \eps$ in constant time (i.e., $\operatorname{poly}(1/\eps)$). Their work, which we draw inspiration from, relies on the ability to sample items from the instance with probability proportional to their profit.

\section{Preliminaries}
    \label{sec:preliminaries}
Throughout the paper, we use $[n]$ to denote the set of integers $\{1,2\,\dots, n\}$, and standard asymptotic notation $O(\cdot),\Omega(\cdot), \Theta(\cdot)$, as well as the (slightly) less standard $\tilde{O}(\cdot)$ which omits polylogarithmic factors in its argument. We recall the definition of $\log^\ast$, the iterated logarithm:
\[
\log^\ast n = \begin{cases}
    0& \text{ if } n \leq 1\\
    1+\log^\ast \log n& \text{ otherwise }
\end{cases}
\]
As out work will be concerned with approximation algorithms, we recall the general definition:
\begin{definition}[$(\alpha, \beta)$-approximation algorithm]\label{def:alphabetaapprox}
    For any $\alpha \in [0,1]$ and $\beta \geq 0$, an $(\alpha, \beta)$-approximation algorithm for an optimization problem that for all instances of the problems produces a solution whose value is
    \begin{enumerate}
        \item at least $(\alpha \cdot \OPT -\beta)$ if the problem is a maximization problem;
        \item at most $(\alpha \cdot \OPT +\beta)$ if the problem is a minimization problem
    \end{enumerate}
    where \OPT is the value of the optimal solution.
\end{definition}

We now formally define Local Computation Algorithms. This definition is the standard definition of~\cite{rubinfeld2011fast}, tailored to the Knapsack problem.\footnote{The usual definition of LCAs includes a first parameter, $s(n)$, for the space complexity of the LCA; we omit it for simplicity, as it is not the focus our of work (neither lower nor upper bounds), which is the query complexity.}
\begin{definition}[LCA for Knapsack Problem]
A \emph{($t(n), \delta (n)$)-LCA $\mathcal{A}$ for Knapsack} is a (randomized) algorithm which is given access to a read-only random seed $r\in\{0,1\}^\ast$, and query access to a Knapsack Instance $I = (S, K)$ on $n=|S|$ items, where the total profit %
of items in $S$ is normalized to 1, and the (integer) weight of any item in $S$ is at most $K$. The algorithm must support the following type of queries: on input $i\in[n]$, after making queries to the instance $I$ and running in time at most $t(n)$, $\mathcal{A}$ outputs whether item $i$ is part of a feasible solution $C$, and must be correct with probability at least $1-\delta(n)$. Importantly, $C$ only depends on the input instance $I$ and random bits $r$ used during computation. Additionally, each run has no access to previous computation results. 
The quantity $t(n)$ is referred to as the time complexity, and $\delta (n)$ the failure probability.\footnote{Note that one would want to set $\delta(n)$ to be $1/\text{poly}(n)$, or at most $O(1/q)$ when the LCA is expected to answer $q$ queries (so that all queries are answered successfully with high probability, by a union bound). Our lower bounds hold even for $\delta(n) = \Omega(1)$, making them even stronger.}
\end{definition}
We also recall two desirable properties of LCAs, which are often taken to be part of the definition:
\begin{definition}[Parallelizable]
    We say an LCA $\mathcal{A}$ is \emph{parallelizable} if multiple copies of $\mathcal{A}$ can be run at the same time. Runs are consistent to the same solution as long as their input Knapsack instance $I$ is the same and use the same random seed $r$ during computation.
\end{definition}

\begin{definition}[Query-Order Oblivious]
    We say an LCA $\mathcal{A}$ is \emph{query-order oblivious} if the outputs of $\mathcal{A}$ do not depend on the order of the queries but depend only on the input and the random seed $r$.
\end{definition}
In what follows, when discussing LCAs we will always refer to parallelizable, query-order oblivious LCAs. Of course, this definition by itself is not enough, as one could achieve a $(O(1), O(1), 1)$-LCA trivially by always answering ``no'' to any query (which would be consistent with the feasible solution $\emptyset$). We typically will require a guarantee on the \emph{profit} of the solution as well.

We will also require the definition of \emph{reproducible algorithm}, as introduced in~\cite{impagliazzo2022reproducibility}. A reproducible algorithm returns the same output on two distinct runs with high probability, provided that (1)~the input samples it takes in both runs come from the same distribution, and (2)~it uses the same internal randomness on both runs:
\begin{definition}[Reproducibility, \cite{impagliazzo2022reproducibility}]
 Let $D$ be a probability distribution over a universe $\mathcal{X}$, and let $\mathcal{A}$ be a randomized algorithm with sample access to $D$. An algorithm $\mathcal{A}$ is \emph{$\rho$-reproducible}
        if
        \[
        \Pr_{{\vec s_1},{\vec s_2},r} \left[
        \mathcal{A}({\vec s_1}; r) = \mathcal{A}({\vec s_2}; r) 
        \right]
        \ge 1 - \rho,
        \]
        where ${\vec s_1}$ and ${\vec s_2}$ denote sequences of samples drawn i.i.d. from $D$, and $r$ denotes a random binary string representing the internal randomness used by $\mathcal{A}$.
   
\end{definition}
In their work, Impagliazzo et al. provide a reproducible algorithm to compute an \emph{approximate median}, defined as follows:
\begin{definition}[$\tau$-approximate median] Let $D$ be a distribution over a well-ordered domain $\mathcal{X}$. For $\tau \in [0,1/2]$, an element $x\in \mathcal{X}$ is
a \emph{$\tau$-approximate median of $D$} if $\Pr_{X\sim D}[X \leq x] \geq 1/2-\tau$ and $\Pr_{X\sim D}[X \geq x] \geq 1/2-\tau$.
\end{definition}
\noindent The guarantees of their algorithm are given in the theorem below:
\begin{theorem}[{\cite[Theorem~4.2]{impagliazzo2022reproducibility}}]
    \label{theo:reproducible:median}
                Let $\tau, \rho \in [0,1]$ and let $\delta = 1/3$. Let $D$ be a distribution over $\mathcal{X}$, where $|\mathcal{X}| = 2^d$. Then there exists an algorithm $\rMedian_{\rho, d, \tau, \delta}$  which is $\rho$-reproducible, outputs a $\tau$-approximate median of $D$ with success probability $1-\rho/2$, and has sample complexity
        \[
        \tilde{\Omega}\left( \left( \frac{1}{\tau^2\rho^2}\right) \cdot \left(\frac{3}{\tau^2}\right)^{\log^{*}|\mathcal{X}|} \right)\,.
        \]
\end{theorem}

Finally, recall that in the Knapsack problem, an instance $I$ consists of a list of $n$ items $(a_1,\dots, a_n)$, each with their own non-negative value $p_i$ and weight $w_i\geq 0$ (so that $a_i = (p_i,w_i)$), and the weight limit $K \geq 0$. The task to select a set $S\subseteq [n]$ of items which maximizes total value $\sum_{i\in S} p_i$ without exceeding the weight limit: $\sum_{i\in S} w_i \leq K$. 

\section{Impossibility Results}

To establish our lower bounds, we will reduce from a \emph{query complexity} problem. Recall that in query complexity, given a function $f\colon\{0,1\}^n\to \{0,1\}$, the objective is to analyse the minimum number of queries (bits) to an input $x\in\{0,1\}^n$ an algorithm has to read in order to successfully compute the value $f(x)$. The \emph{randomised query complexity} $R(f)$ is the worst-case query complexity (over all possible inputs $x\in\{0,1\}^n$) of any randomised algorithm which correctly computes $f$, on any input, with probability at least $2/3$. For more on query complexity, the reader is referred to the survey~\cite{BuhrmanW02}. In particular, we will use the following standard result on the OR function, $\operatorname{OR}_n$, defined as $\operatorname{OR}_n(x) = \bigvee_{i=1}^n x_i$:
\begin{lemma}
    \label{lemma:orn:querycomplexity}
    The randomised query complexity of the OR function on $n$ bits is $R(\operatorname{OR}_n) = \Omega(n)$.
\end{lemma}
\begin{theorem}
    \label{theo:no:optimal:knapsack:lca}
    Any $(t(n),1/3)$-LCA for Knapsack which provides query access to an optimal solution must satisfy $t(n) = \Omega(n)$.
\end{theorem}

\begin{proof}
Let $\mathcal{A}$ be any $(t(n),1/3)$-LCA for Knapsack. We will show how to use it to compute the function $\operatorname{OR}_{n-1}$ with at most $t(n)$ queries to the input (and success probability at least $2/3$: by~\cref{lemma:orn:querycomplexity}, this will imply $t(n)=\Omega(n)$.

On input $x\in\{0,1\}^{n-1}$, we will simulate access the Knapsack instance $I(x)$ on $n$ items and weight limit $K=1$ defined as follows:
\[
(p_i, w_i) =\begin{cases} (x_i,1) &\text{ if } 1\leq i\leq n-1 \\
(1/2,1) &\text{ if } i=n
\end{cases}
\]That is, all $n$ items have weight equal to the weight limit, and so any feasible solution can only contain at most one item. It is easy to see that $\operatorname{OR}_{n-1}(x)=1$ if, and only if, an optimal solution to the Knapsack instance has value $1$; otherwise, it has value $1/2$ (as $x_i = 0$ for all $1\leq i\leq n-1)$), and the only optimal solution is the singleton $\{s_{n}\}$ which has profit $1/2$.

\begin{figure}[htbp]\centering
\begin{tikzpicture}[scale=1.5]\scriptsize
    \foreach \y in {1,2,3,4} {
        \node[draw,fill=black!10!white,minimum size=5em] at (\y,0) (block\y) {$x_{\y}$};
    }
    \node[minimum size=5em] at (5,0) (blockdots){$\cdots$};
    \node[draw,fill=black!10!white,minimum size=5em] at (6,0) (blockn2){$x_{n-2}$};
    \node[draw,fill=black!10!white,minimum size=5em] at (7,0) (blockn1){$x_{n-1}$};

    \foreach \y in {1,2,3,4} {
        \node[draw,minimum size=5em] at (\y,-1.5) (block\y) {$(x_{\y},1)$};
    }
    \node[minimum size=5em] at (5,-1.5) (blockdots){$\cdots$};
    \node[draw,minimum size=5em] at (6,-1.5) (blockn2){$(x_{n-2},1)$};
    \node[draw,minimum size=5em] at (7,-1.5) (blockn1){$(x_{n-1},1)$};
    \node[draw,minimum size=5em] at (8,-1.5) (blockn){$(\frac{1}{2},1)$};
\end{tikzpicture}
    \caption{An illustration of the reduction. Given query access to an input $x\in\{0,1\}^{n-1}$ (top), we simulate query access to a Knapsack instance $I(x)$ (bottom) with weight limit $K=1$, such that the $n$-th item belongs to the optimal solution (which is then unique) if, and only if, $\operatorname{OR}_n(x)=0$.}
\end{figure}
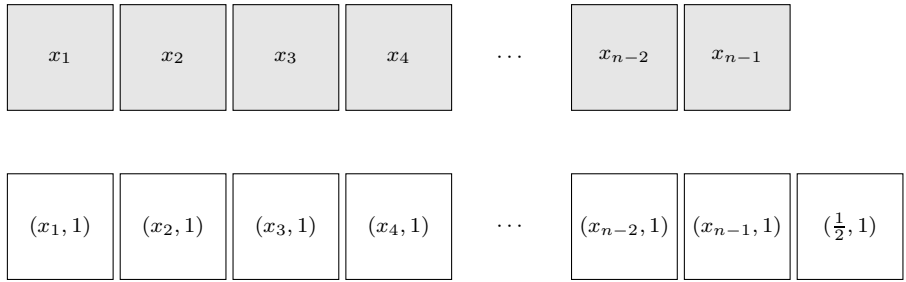
Moreover, one can simulate any query made by $\mathcal{A}$ to this instance $I(x)$ at the cost of (at most) one query to $x$: 
\begin{itemize}
    \item If $\mathcal{A}$ queries item $s_{n}$, answer $(1/2,1)$ (no query to $x$ needed);
    \item If $\mathcal{A}$ queries item $s_{i}$ for some $1\leq i\leq n-1$, query $x_i$ and answer $(x_i,1)$ (one query to $x$ needed).
\end{itemize}
Given the above, to compute $\operatorname{OR}_n(x)$, it suffices to make a \emph{single} query to the LCA $\mathcal{A}$: namely, the query asking if $s_{n}$ is in the optimal solution it gives access to. By the above discussion, $s_{n}$ is part of an optimal solution if, and only if, $\operatorname{OR}_{n-1}(x)=0$, so the answer given by $\mathcal{A}$ on this single query provides (with probability at least $2/3$) the value $\operatorname{OR}_{n-1}(x)$.

To conclude, note since the LCA $\mathcal{A}$ can answer any single query to an optimal solution to our instance $I(x)$ (of size $n$) in time at most $t(n)$, it makes at most $t(n)$ queries to the instance $I(x)$: indeed, each query takes unit time, so the number of queries is a lower bound on the time complexity. Overall,
\[
t(n) \geq \substack{\text{number of queries by }\mathcal{A}\\\text{to the instance }I(x)} \geq \substack{\text{number of queries}\\\text{made to the input }x}
\]
So our reduction allows us, on any input $x$, to compute (with probability at least $2/3$) $\operatorname{OR}_{n-1}(x)$ with at most $t(n)$ queries to $x$: by~\cref{lemma:orn:querycomplexity}, it follows that $t(n) = \Omega(n)$.
\end{proof}
The above result rules out LCAs which provide local access to an optimal Knapsack solution. We will reduce from the same query complexity problem to a slightly different Knapsack Instance to show that any LCA that gives query access to \emph{any} finite approximate Knapsack solution takes at least $\Omega(n)$ time.
\begin{theorem}
    \label{theo:no:approx:knapsack:lca}
    Fix any $\alpha\in(0,1]$. Any $(t(n),1/3)$-LCA for Knapsack which provides query access to an $\alpha$-approximate feasible solution must satisfy $t(n) = \Omega(n)$.
\end{theorem}
\begin{proof}
    Let $\mathcal{A}$ be any $(t(n),1/3)$-LCA for $\alpha$-approximation Knapsack, and fix any $0<\beta<\alpha$. On input $x \in \{0,1\}^{n-1}$, we simulate access the Knapsack Instance $I(x)$ on $n$ items and weight limit $K = 1$ defined as follows:
    \[
    (p_i, w_i) =\begin{cases} (x_i,1) &\text{ if } 1\leq i\leq n-1 \\
(\beta,1) &\text{ if } i=n
\end{cases}      \]
i.e., the same instance as in the proof of~\cref{theo:no:optimal:knapsack:lca}, but with the profit of the last item set to $\beta$. As before, any feasible solution of $I(x)$ contains at most one item as all $n$ items have the weight equal to the weight limit. The value of the optimal solution is then either $1$ (if at least one of the the $x_i$'s is $1$) or $\beta$ (if all $x_i$'s are $0$, in which case $\{s_n\}$ is the optimal solution. Thus $\{s_n\}$ is an $\alpha$-approximation solution if, and only if, $\operatorname{OR}_{n-1}(x)=0$; in this case, $s_n=(\beta, 1)$ has higher profit than any other items which makes it the optimal solution, hence it is also a unique $\alpha$-approximation solution for $I(x)$. Otherwise the optimal solution contains an item with profit $1$, and $\{s_n\}$ is not an $\alpha$-approximation solution as $\beta < 1 \cdot \alpha$. Again, we can simulate any query by $\mathcal{A}$ to this instance $I(x)$ at the cost of one query to $x$:
\begin{itemize}
    \item If $\mathcal{A}$ queries item $s_{n}$, answer $(\beta,1)$ (no query to $x$ needed);
    \item If $\mathcal{A}$ queries item $s_{i}$ for some $1\leq i\leq n-1$, query $x_i$ and answer $(x_i,1)$ (one query to $x$ needed).
\end{itemize}
It is sufficient to compute $\operatorname{OR}_{n-1}(x)$ by making a single query to the LCA $\mathcal{A}$: by asking whether $s_n$ is in a $\alpha$-approximation solution it gives access to. Since the LCA $\mathcal{A}$ can answer any single query to a $\alpha$-approximation solution to our instance $I(x)$ (of size $n$) in at most $t(n)$ time, $\mathcal{A}$ makes at most $t(n)$ queries to the instance $I(x)$. And any query made to $I(x)$ can by simulated by at most one query to $x$. Therefore, $t(n)$ is lower bounded by the query complexity of $\operatorname{OR}_{n-1}(x)$. By~\cref{lemma:orn:querycomplexity}, it follows that $t(n) = \Omega(n)$.
\end{proof}
Since we can choose $\alpha$ to be arbitrarily close to 0, we get that there is no sublinear time LCA that answers to any finite approximation Knapsack solution with failure probability at most 1/3.

Having ruled out in~\cref{theo:no:optimal:knapsack:lca,theo:no:approx:knapsack:lca} any sublinear-time LCAs for optimal and approximate Knapsack, we further relax the constraint to show that even asking for an different relaxation of Knapsack is impossible. That is, instead of giving answers consistent to a finite approximation solution, we ask if an LCA can give access to a \emph{maximal feasible} Knapsack solution in sublinear time. Instead of proving the impossibility result directly for LCAs, we will show that no \emph{deterministic} algorithm can provide query access to a maximal feasible solution on some difficult input distribution in sublinear time. By Yao's Principle~\cite{4567946}, this will imply that no randomized algorithm can provide sublinear time query access to maximal feasible Knapsack. 
\begin{theorem}
    \label{theo:no:maximal:knapsack:lca}
    Any $(t(n),4/5)$-LCA for Knapsack which provides query access to an maximal feasible solution must satisfy $t(n) = \Omega(n)$.
\end{theorem}
\begin{proof}
We define the distribution of items $S$ of the input Knapsack instance $I$ as follows:
\begin{enumerate}
    \item select uniformly at random a pair of indices $(i, j) \in \{1, 2, \dots, n\}^2$
    \item assign $w_i = 3/4$, and 
    \[
        w_j = \begin{cases}
            1/4 &\text{ with probability } 1/2\\
            3/4 &\text{ with probability } 1/2
        \end{cases}
    \]
    \item every other item $k \notin \{i,j\}$ is assigned $w_k = 0$.
\end{enumerate}
Let $\mathcal{A}$ be a \emph{deterministic} algorithm for Maximal Feasible Knapsack with time complexity $t(n) < \frac{n}{11}$ and probability of success at least $4/5$.

The weight limit $K$ is 1. (Since we are looking at maximal feasible solutions, the profits of the items are irrelevant: we set them all to $0$ (i.e., set $p_k=0$ for all $k$) and ignore them for the rest of the proof.) 
Note that if $w_j=1/4$, then the (unique) maximal solution is $C = \{s_1,\dots,s_n\}$ (all items); while if $w_j=3/4$, then the two distinct maximal solutions are $C^{(-i)} = \{s_1,\dots,s_n\}\setminus\{s_i\}$ and $C^{(-j)} = \{s_1,\dots,s_n\}\setminus\{s_j\}$. The following claim will be crucial:
\begin{lemma}
    Suppose $\mathcal{A}$ receives query $s_k$, such that $w_k=3/4$. Then with probability at least $9/10$ over the choice of the instance $I$, $\mathcal{A}$ outputs \textsf{yes} (i.e., that this item is in the maximal feasible solution).
\end{lemma}
\begin{proof}
    Given such a query $s_k$, in view of our choice of input distribution, there are two options: either $k=i$ (since $w_i=3/4$ always), or $k=j$ and $w_j=3/4$. So $k\in\{i,j\}$: let $k'$ be the other value of the couple (that if, if $k=i$ then $k'=j$, and vice-versa).

    On query $s_k$, we assume that the algorithm knows that $w_k=3/4$ (since this only take one query to the instance $I$ to learn this, given $k$).    
    The algorithm makes deterministically a sequence of at most $q\leq t(n) < \frac{n}{11}$ queries to the instance $I$, where query $\ell$ reveals the weight $w_\ell$ of item $\ell$. Without loss of generality, we can assume that the algorithm does not query an item it already knows the weight of; so it queries at most $q < \frac{n}{11}$ new items. Since $k'$ is uniformly distributed among the remaining $n-1$ items, the probability that any of these $q$ queries (even made adaptively) finds the other non-zero-weight item $s_{k'}$ is at most
\[
    \frac{n}{n-1}\cdot \frac{1}{11} \leq \frac{1}{10}
\]
since the answer to every single query $\ell$ made is answered by ``$w_\ell=0$.'' We denote by $E_i$ the event
\[
    E_k = \{\text{on query }k,\, A \text{ does not query any other item with non-zero weight }\}
\]
By the above, $\Pr[E_k] \geq \frac{9}{10}$. Now, the claim is that conditioned on $E_k$, the algorithm must say \textsf{yes} to the query (i.e., that $s_k$ is in the maximal solution it provides access to). To see why, note that given only the information given (namely, the fact that the $k$-th item has weight $3/4$, and that all the other items queried have weight $0$), the three following cases are possible given our distribution over instances $I$:
\begin{itemize}
    \item the $k$-th item is $s_j$, and there is somewhere another item, $s_i$, with weight $3/4$ (this has probability $1/3$)
    \item the $k$-th item is $s_i$, and there is somewhere another item, $s_j$, with weight $3/4$ (this has probability $1/3$)
    \item the $k$-th item is $s_i$, and there is somewhere another item, $s_j$, with weight $1/4$ (this has probability $1/3$)
 \end{itemize}
In the first two cases, the algorithm could reply either \textsf{yes} or \textsf{no} and still be consistent with one of the two maximal solutions; but in the third case, the algorithm must respond \textsf{yes} since \emph{all} items are in the (unique) maximal solution. Since we are in the third case with probability $1/3$ (and that the algorithm gets no useful information except with probability at most $1/10$), to achieve error less than $1/5 < 1/3-1/10$ overall the algorithm must respond \textsf{yes}.
\end{proof}
By the \emph{principle of deferred decisions}, we can make the random choice of deciding the value of $w_j$ (either $1/4$ or $3/4$ uniformly at random) when $\mathcal{A}$ queries $s_j$.

Consider what happens when the LCA $\mathcal{A}$ receives queries for item $s_i$, then $s_j$. By the above claim (for $k=i$), on query $s_i$, $\mathcal{A}$ will output \textsf{yes} with probability at least $9/10$ (over the choice of the instance).

On the \emph{second} query $s_j$, we then have two choices: with probability $1/2$, $w_j=1/4$ and the algorithm can safely respond $\textsf{yes}$. With probability $1/2$, however, $w_j=3/4$, and by the same claim (for $k=j$) will then output \textsf{yes} with probability at least $9/10$.

This means, by a union bound, that the algorithm is correct with probability at most
\[
\Pr[\bar{E_i}] + \Pr[w_j \neq  \frac{3}{4}] +  \Pr[\bar{E_j}]
\leq \frac{1}{10} + \frac{1}{2}+ \frac{1}{10} < \frac{4}{5}
\]
since with probability $\Pr[E_i\cap \{w_j=3/4\} \cap E_j]$ it says \textsf{yes} to item $s_i$ with weight $3/4$ \emph{and} item $s_j$ which has weight $3/4$; i.e., the solution it is consistent with is not feasible.

This implies that any deterministic algorithm for maximal feasible Knapsack which is correct with probability at least $4/5$ on this specific input distribution must make at least $\frac{n}{11}$ queries. By Yao's Principle, this implies that any randomized algorithm cannot answer Maximal Feasible Knapsack with success probability $4/5$ unless $t(n) \geq \frac{n}{11}$.
\end{proof}
\section{Algorithmic Result: an LCA given Weighted Sampling Access}
We have shown in the last section that it is impossible to obtain a (non-trivial) LCA for Knapsack, or even some relaxations of it. The crux of the issue lies in the ``needle in a haystack'' phenomenon: in order to be consistent across queries, an LCA seemingly needs to get a global sense of the distribution of weights and profits of the instance, and in particular the largest profits. Yet, given only query access to the instance, finding these requires a very large number of queries.

To mitigate this, we adapt the weighted sampling model considered in \cite{ito2012constant}, which equips the algorithm with the ability to sample items with probability proportional to their profit (assuming, essentially without loss of generality, that the total profit and weight are both normalized to 1). In this setting, we establish the following result:
\begin{theorem}\label{theo:lcakpresult}
    \cref{alg:KP_LCA} ($\lcakp$) is a $(t(n),\eps)$-LCA for Knapsack which provides query access to an $(1/2, 6\eps)$-approximate solution, where $t(n) = ( 1/\eps )^{O(\log^{*}n)}$.
\end{theorem}
The remainder of this section is dedicated to proving this theorem. 
Before we describe the general idea of the algorithm, we first cover the weighted sampling model and results from \cite{ito2012constant}  we will use in our algorithm. Fixing $\eps \in (0,1]$, the items of the Knapsack instance $I = (S, K)$ are partitioned into three sets: 
        \begin{align*}
            L(I) &:= \{ (p,w) \in S : p > \eps^2\} &\text{ (items with high profit) }\\
            S(I) &:= \{ (p,w) \in S : p \leq \eps^2 \text{ and } p/w \geq \eps^{2}\}  &\text{ (items with low profit but high efficiency) }\\
            G(I) &:= \{ (p,w) \in S : p \leq \eps^2 \text{ and } p/w < \eps^{2}\}  &\text{ (items with low profit and low efficiency) } 
        \end{align*}
The items in $L(I)$, $S(I)$, $G(I)$ are referred as \emph{large items}, \emph{small items} and \emph{garbage items} respectively. The following lemma, which follows from a coupon-collector type argument, ensures that one can sample all sufficiently heavy elements with high probability. 
\begin{lemma}[Lemma 2, \cite{ito2012constant}]\label{lem:large:item:weighted:sampling}
Let $B = \{(w,p) \in I : p \geq \delta \}$. By taking $\lceil 6\delta^{-1}(\log\delta^{-1} + 1)\rceil$ samples using weighted sampling, all items of $B$ are sampled at least once with probability at least~$5/6$.
\end{lemma}
Importantly, this success probability can be easily amplified by repetition from $5/6$ to $1-\eps$ at the cost of a logarithmic factor in $1/\eps$, which we will rely on later.

Small items are partitioned over intervals of efficiency such that the total profits of items within each interval is approximately $\eps$. The set of small items in each efficiency intervals is denoted by  $A_{0}(I), A_{1}(I),..., A_{t}(I)$, and the corresponding \emph{efficiency thresholds} are denoted by a non-increasing efficiency sequence $e_1 \geq e_2 \geq ... \geq e_t$, such that:
\begin{align*}
    A_{0}(I) &:= \{ (p,w) \in S(I) : p/w \geq e_1\}\\
    A_{k}(I) &:= \{ (p,w) \in S(I) : e_k > p/w \geq e_{k+1} \} \tag{$1 \leq k \leq t-1$}\\
    A_{t}(I) &:= \{ (p,w) \in S(I) : e_t > p/w\}
\end{align*}

A crucial definition is then that of an \emph{equally partitioning sequence}, which is a sequence of efficiency threshold such that the total profit of each efficiency interval satisfies $\sum_{i\in A_k(I)} p_i \approx \eps$:
\begin{definition}[Equally Partitioning Sequence (EPS), \cite{ito2012constant}]
    We say the efficiency sequence $e_1, e_2, ..., e_t$ is \emph{equally partitioning with respect to $I$} if, for all $0\leq i \leq t-1$, $A_{i}(I)$ has total profit within $[\eps, \eps + \eps^{2})$ and $A_t(I)$ has total profit within  $[0, \eps + \eps^{2})$. 
\end{definition}
The authors of~\cite{ito2012constant} also show how to obtain an EPS, with probability at least $1-O(1/m^2)$, from sampling $O(\frac{1}{\eps^4}\log\frac{1}{\eps})$ items.\footnote{The statement in~\cite{ito2012constant} only claims success probability $5/6$; however, inspection of the proof show the authors establish the stronger claim.} Now, their algorithm, given weighted sampling access to the original instance $I$ and a parameter $\eps$, constructs a different instance $\tilde{I}$ (of size $\operatorname{poly}(1/\eps)$) as follows:
\begin{enumerate}
    \item It samples a multiset of $O(\frac{1}{\eps^4}\log\frac{1}{\eps})$ items, and puts all the large items into a set $M$. (By~\cref{lem:large:item:weighted:sampling}, all items of $L(I)$ are included in $M$, except with probability $1/6$.)
    \item It samples a multiset of $O(\frac{1}{\eps^4}\log\frac{1}{\eps})$ items, and uses it to obtain an Equally Partitioning Sequence $\tilde{e}_1, \dots, \tilde{e}_t$, except with probability $1/6$.
    \item\label{construction} From the above set $M$ and the EPS, it constructs a new instance $\tilde{I} = (\tilde{S},\tilde{K})$ as follows. It creates three disjoint sets $L(\tilde{I}), S(\tilde{I}), G(\tilde{I})$ mimicking the sets $L(I), S(I), G(I)$ defined for $I$:
\begin{align*}
    L(\tilde{I}) &:= M,\\
    S(\tilde{I}) &:= \bigcup_{k = 0}^{t-1}A_{k}(\tilde{I}), \text{ where } A_{k}(\tilde{I}) \text{ contains exactly }\lfloor \eps^{-1}\rfloor \text{ copies of }(\eps^2, \eps^{2}/\tilde{e}_{k+1}),\\
    G(\tilde{I}) &:= \emptyset 
\end{align*}
and defines $\tilde{S} := L(\tilde{I}) \cup  S(\tilde{I}) \cup G(\tilde{I}) $ and $\tilde{K}=K$.
\end{enumerate}
Note that $t=O(1/\eps)$ and $|M| \leq |L(I)| \leq 1/\eps^2$, and so $\tilde{I}$ is an instance on only $O(1/\eps^2)$ items. 
We call the above procedure the \emph{$\tilde{I}$-construction algorithm}.
The central lemma they show then relates the value of the optimal solution to the new Knapsack instance $\tilde{I}$ to that of the original instance $I$.
\begin{lemma}[Lemma 1, \cite{ito2012constant}]\label{lem:relation:tilde:I}
Let $I = (S, K)$ be an instance of Knapsack, and let $\tilde{I}$ be the instance constructed from $I$ using an equally partitioning sequence with respect to $I$. Then, $\OPT(\tilde{I})-\eps$ is a $(1, 6\eps)$-approximation to $\OPT(I)$.
\end{lemma}

\subsection{General Idea of the Algorithm}
  As in the algorithm of~\cite{ito2012constant}, we will use weighted sampling to gather items with high profit and learn the distribution of the efficiencies of small items with high probability. Then we will compute the greedy solution of the corresponding simplified Knapsack instance $\tilde{I}$, as defined above. (Recall that the greedy algorithm for the Knapsack problem first sorts the items over their efficiency, then it includes the items in descending order until the output set reaches the weight limit; we will refer to the efficiency of the first item that the greedy solution cannot fully include as the \emph{efficiency cut-off}.)  The greedy solution will either be the set of items with efficiency higher than the efficiency cut-off, or the one item with efficiency cut-off if that item's profit is higher than the total profit of the items with higher efficiency.
 
 Our LCA will determine the queried item is chosen or not based on the efficiency cut-off of the greedy solution of the simplified Knapsack of that run. As discussed in~\cref{sec:intro:results}, one challenge lies in the fact that the simplified instance $\tilde{I}$ is constructed by sampling, hence items of the simplified instance and the corresponding efficiency cut-off may (and likely will) be different between each run, jeopardizing the consistency requirement of the LCA. To address this issue, we generalize the \emph{reproducible median} algorithm from \cite{impagliazzo2022reproducibility} to a \emph{reproducible quantile} algorithm that will output the exact same approximation quantiles of the efficiency of the items with high probability. We will start with the \emph{reproducible quantile} algorithm.

\subsection{Reproducible Quantile Algorithm}
We will leverage the reproducible algorithm of~\cite{impagliazzo2022reproducibility} (\cref{theo:reproducible:median}) to compute the quantiles in a reproducible fashion. 
 The idea is to reduce the task of computing quantiles to that of computing the median. Given an array $T$ of $n$ elements, to find the $p$-quantile of $T$, we can add $x$ many $-\infty$ and $y$ many $+\infty$ to $T$, such that $x + pn = (1-p)n + y$ and $x + y = n$, let the new array be $T'$. The median of array $T'$ will be the same value as the $p$-quantile of array $T$.

Similarly, we can compute the reproducible $p$-quantile of a given distribution $D$ by computing the reproducible median of a new distribution $D'$ such that:
\begin{enumerate}
    \item The domain of $D'$ is the union of the domain of $D$ (assumed to be of size $2^d$) and $\{+\infty, -\infty \}$; note that we can bound domain size of $D'$ by $2^{d+1}$,
    \item For any element $e$ from distribution $D$, $\Pr_{i \sim D'}[i = e] = \frac{\Pr_{i \sim D}[i = e]}{2}$,
    \item $\Pr_{i \sim D'}[i = -\infty] = x, \Pr_{i \sim D'}[i = +\infty] = y, x = (1-p)/2, y = p/2$.
\end{enumerate}

\begin{algorithm}[htbp]
	\begin{algorithmic}[1]
    \Require $\vec s$: a sample of $n_d$ elements drawn i.i.d. from $D'$; $p$: queried percentile in $D$ 
	\Require Parameters: 
        \begin{itemize}
			\item $\rho$: target reproducibility parameter 
			\item $d$: specifies domain size $| \mathcal{X} | = 2^d$ 
			\item $\tau$: target accuracy of $p$-quantile
			\item $\beta$: target failure probability
            \item $n_{d}$: sample complexity of $\rMedian$ to output a $\rho$-reproducible $\tau$-approximation median with failure probability $\beta$ on domain $|\mathcal{X}| = 2^{d}$ 
        \end{itemize}
	\Ensure returns a $\tau$-approximate $p$-quantile of $D$
	   \State Run $\rMedian(\vec s)$ with reproducibility parameter $\rho$, domain size $2^{d+1}$, target accuracy $\tau/2$ and target failure probability $\beta$ and return the result $v$
	\end{algorithmic}
	\caption{$\rQuantile_{}(\vec s, p)$}\label{algo:rquantile}
\end{algorithm}

\begin{theorem}[$\rQuantile$ Correctness and Query Complexity]
    \label{theo:rquantile}
    Let $\tau, \beta, \rho \in [0,1]$. Let $D$ be a distribution over $\mathcal{X}$, with $|\mathcal{X}| = 2^d$. Then
	$\rQuantile_{\rho, d, \tau, \beta}$ (\cref{algo:rquantile}) is a reproducible algorithm with sample complexity 
		\[
        \tilde{O}\left( \left( \frac{1}{\tau^2(\rho - \beta)^2}\right) \cdot \left(\frac{12}{\tau^2}\right)^{\log^{*}|\mathcal{X}| + 1} \right)
        \]
        which, on input $p$, outputs a $\tau$-approximate $p$-quantile of $D$ with probability at least $1-\beta$.
\end{theorem}
\begin{proof}
    We first argue the correctness of $\rQuantile$. $\rQuantile$ returns a value $v$ which is a $\rho$-reproducible $\tau/2$-approximation median of distribution $D'$ except with probability $\beta$. Thus,
    \[
    \Pr_{i \sim D'}[i \leq v] \geq 1/2 - \tau/2 \text{ and } \Pr_{i \sim D'}[i \geq v] \geq 1/2 - \tau/2.
    \]
    By the definition of $D'$,
    \[
        \Pr_{i \sim D}[i \leq v] = 2\Pr_{i \sim D'}[i \leq v \text{ and } i \neq -\infty] \geq 2 (1/2-\tau/2 - (1-p)/2) = p - \tau
        \]
        and
  \[
        \Pr_{i \sim D}[i \geq v] = 2\Pr_{i \sim D'}[i \geq v \text{ and } i \neq +\infty] \geq 2 (1/2-\tau/2 - p/2) = 1 - p - \tau.
        \]
    Therefore, $\rQuantile$ returns a $\tau$-approximation $p$-quantile of distribution $D$.
    By~\cite[Theorem 4.8]{impagliazzo2022reproducibility}, $\rMedian_{\rho, d, \tau, \beta}$ has sample complexity of 
    \[\tilde{O}\left( \left( \frac{1}{\tau^2(\rho - \beta)^2}\right) \cdot \left(\frac{3}{\tau^2}\right)^{\log^{*}|\mathcal{X}|} \right).
    \]
    Since $\rQuantile$ runs $\rMedian$ with target accuracy $\tau/2$ and domain size $2^{d+1}$, the sample complexity of $\rQuantile$ is 
    \[
    \tilde{O}\left( \left( \frac{1}{\tau^2(\rho - \beta)^2}\right) \cdot \left(\frac{12}{\tau^2}\right)^{\log^{*}|\mathcal{X}| + 1} \right)\,,
    \]
    as desired.
\end{proof}

\paragraph{Mapping to a finite domain.} $\rQuantile$ can only compute quantiles with respect to a distribution $D$ over a \emph{finite} domain $\mathcal{X}$ (of some size of the form $2^d$) since the recursion of $\rMedian$ terminates when the domain size become 2. However, in our case the distribution $D$ is the distribution of efficiencies, which (since both profits and weights are \emph{a priori} arbitrary positive numbers) have domain $\mathbb{R}_{>0}$. To address this, we first note that the weights are not quite arbitrary: before the normalization assumption, they were all positive integers. 
Assuming a $\operatorname{poly}(n)$ bound on the number of bits needed to describe any given weight, we get that the domain of the (normalized) weights is of the form $\mathcal{W} = \{1/B, 2/B, \dots, 1 \}$, where $B = 1/2^{\operatorname{poly}(n)}$.

We can assume the same for the profits, i.e., that all profits take value in the domain $\mathcal{P} = \{0,1/B', 2/B', \dots, 1 \}$ for some $B' = 1/2^{\operatorname{poly}(n)}$.\footnote{Another possibility is to avoid this bit complexity argument and instead use a rounding procedure, standard in the fully polynomial-time approximation scheme (FPTAS) for Knapsack (see~\cite[Section~3.2]{williamson2011design}), which rounds each profit ``on-the-fly'' to a multiple of $\varepsilon/n$, as the price of an $1-\varepsilon$ approximation factor loss in the solution. This would bound the size of the domain of the profits by $n/\varepsilon$, for any fixed choice of $\varepsilon>0$. We here describe the bit-complexity bound argument, as it mimics that of the weights.} Overall, this ensures that each efficiency is of the form 
\[
    \frac{B}{B'} \cdot \frac{a}{b}
\]
for some $0\leq a \leq B'$ and $1\leq b \leq B$. This leads to the domain $\mathcal{X}$ of the efficiencies being known and finite (but huge), of size bounded by $|\mathcal{X}| = (B'+1)B = 2^{\operatorname{poly}(n)}$. Note that this implies that the dependence on the domain size from $\rQuantile$, which involves $\log^\ast |\mathcal{X}|$ (in the exponent), will be sublinear in $n$, as
\[
\log^\ast|\mathcal{X}| = \log^\ast(2^{\operatorname{poly}(n)}) = O(\log^\ast n).
\]

\subsection{From Reproducible Quantiles to our Knapsack LCA}
We now put all the pieces together to analyze our final algorithm,~\cref{alg:KP_LCA}, thus establishing~\cref{theo:lcakpresult}. We will show that the solution $C$ the algorithm answers according to is (1)~feasible (\cref{lem:mapping:feasible}), (2)~a good approximate solution (\cref{lem:mapping_approx}), and (3)~consistent across queries (\cref{lem:lcakpconsistency}); before bounding the sample (query) complexity in~\cref{lem:lcakpsample}.

\begin{algorithm}[htbp]
\caption{$\lcakp$}\label{alg:KP_LCA}
\begin{algorithmic}[1]\small
    \Require $I = (S, K)$: Knapsack Instance, where $S$ is the item set and $K$ is the weight limit; $i$: queried index of item in $S$
\State\label{l:get:R} Let $\vec R$ be a sample of size  $m := O(\frac{1}{\eps^4}\log \frac{1}{\eps})$ %
from $S$ \Comment{Succeeds with probability $1-\eps/3$ by \cref{lem:large:item:weighted:sampling}}
\State Remove all items with profit less or equal than $\eps^{2}$ and duplicate items from $\vec R$ 
\State$L(\tilde{I}) \gets \vec R$
\If{$1 - p(L(\tilde{I})) \geq \eps$} \label{line:checkifepspossible}
\State Set $q := \frac{\eps + \eps^{2}/2}{1 - p(L(\tilde{I}))}$, $t = \lfloor q^{-1} \rfloor, \tau = \eps^{2}/5, \rho = \frac{\eps^2}{18}, \beta = \rho/2,  n_{rq} = $ sample complexity of $\rQuantile_{\rho, d, \tau, \beta}$
\State Let $\vec Q$ be a sample of size $a :=  \lceil \frac{3n_{rq}}{2(1-p(L(\tilde{I}))) } \rceil$ from $I$ \label{line:samplesize}
\State Remove every item with profit greater than $\eps^2$ from $\vec Q$
\State $\vec E \gets \{p/w \mid (p,w) \in \vec Q\}$
\For{ $k = 1, 2, \dots, t$ }
\State $\tilde{e}_k \gets \rQuantile_{\rho, d, \tau, \beta}(\vec E, 1-kq)$
\EndFor
\If{$\tilde{e}_t < \eps^2$} \label{line:checkt}
\State $t' \gets t - 1$
\Else \State $t' \gets t$
\EndIf
\State $\text{EPS} \gets (\tilde{e}_1, \tilde{e}_2, \dots, \tilde{e}_{t'})$
\Else \State $\text{EPS} \gets \emptyset$ \label{line:emptyeps}
\EndIf
\State Construct Knapsack Instance $\tilde{I} = (\tilde{S}, K)$ based on $L(\tilde{I})$ and efficiency sequence $\text{EPS}$ as in step~\ref{construction} of the $\tilde{I}$-construction algorithm
\State $\text{Index}_{\rm{}large}, e_{\rm{}small}, B_{\rm{}indicator} \gets \efficiencythreshold(\tilde{I}, \text{EPS} )$
\If{$p_i > \eps^{2}$}
\State return yes if $i \in \text{Index}_{\rm{}large}$, otherwise return no
\ElsIf{$e_{\rm{}small} \neq -1$ and $p_i/w_i \geq e_{\rm{}small} $}
\State return \textsf{yes}
\EndIf
\State return \textsf{no}
\end{algorithmic}
\end{algorithm}
\begin{algorithm}[htbp]
    \caption{$\efficiencythreshold$}\label{alg:new_efficiency_threshold}
    \begin{algorithmic}[1]\small
    \Require $\tilde{I}$ and $\tilde{e}_1, \tilde{e}_2, \dots, \tilde{e}_t$, the Equally Partitioning Sequence that $\tilde{I}$ is constructed from 
    \State Let $\{(p_1, w_1), \dots, (p_{|\tilde{S}|}, w_{|\tilde{S}|})\}$ be the items in $\tilde{S}$ sorted by their efficiency in non-increasing order.
    \State Let $1\leq j\leq |\tilde{S}|$ be the largest index such that $\sum_{i=1}^{j}w_i \leq K$
     \State Let $1 \leq k \leq t$ be the largest index such that $\tilde{e}_{k} > p_{j}/w_{j}$
    \If{$j = |\tilde{S}|$ or $\sum_{i=1}^{j}p_i \geq p_{j+1}$}
        \State $\text{Index}_{\rm{}large} \gets \{ \text{the index in $I$ corresponding to } i \mid 1\leq i\leq j,  (p_i,w_i) \in I, p_i > \eps^2$\}
        \If{$k \geq 3$} \State $e_{\rm{}small} \gets e_{k-2}$ \Else \State $e_{\rm{}small} \gets -1$ \EndIf
        \State $B_{\rm{}indicator} \gets \textsf{false}$
    \Else
        \State $\text{Index}_{\rm{}large} \gets \{ \text{the index in $I$ corresponding to } j+1\}$
        \State $ e_{\rm{}small} \gets -1,  B_{\rm{}indicator} \gets \textsf{true}$
    \EndIf
    \State\Return $\text{Index}_{\rm{}large}, e_{\rm{}small}, B_{\rm{}indicator}$
    \end{algorithmic}
\end{algorithm}
\begin{algorithm}
    \caption{$\mappinggreedy$}\label{alg:convertgreedy}
        \begin{algorithmic}[1]\small
    \Require $\text{Index}_{\rm{}large}, e_{\rm{}small}, B_{\rm{}indicator}$: Output of $\efficiencythreshold$; $I$: the original Knapsack Instance 
    \State $C \gets \{(p_{i}, w_{i}) \mid i \in \text{Index}_{\rm{}large}\}$
    \If{$B_{\rm{}indicator} = \textsf{false}$  and $e_{\rm{}small} \neq -1$}
        \State $C \gets C \cup \{(p, w) \mid (p,w)\in S(I) \text{ and } p/w \geq e_{\rm{}small}\}$ 
    \EndIf \\
    \Return{$C$}
    \end{algorithmic}
\end{algorithm}
Without loss of generality, we will refer to the items by their index after sorting in $\efficiencythreshold$. We separate $\efficiencythreshold$ and $\mappinggreedy$ as we will only call $\efficiencythreshold$ in $\lcakp$.
\begin{lemma}\label{lem:EPSfailureprobability}
    Conditioned on the sample $\vec{R}$ containing all elements of $L(I)$  after \cref{l:get:R}, then, with probability at least $1- 13\eps/36$, the approximate quantile sequence $\tilde{e}_1, \dots, \tilde{e}_{t'}$ in $\lcakp$ (\cref{alg:KP_LCA}) is an EPS with respect to $I$.
\end{lemma}
\begin{proof}
    $\rQuantile_{\rho, d, \tau, \beta}$ will return $\tau$-approximate $(1-kq)$-quantiles over the distribution of efficiencies of items drawn i.i.d from $S(I) \cup G(I)$ with probability $1-\beta$ as long as $|\vec E|$ satisfies the required sample complexity. For $t = \lfloor q^{-1} \rfloor, k = 1, 2, \dots, t$, we have:
    \[
        \Pr_{(p,w) \sim S(I) \cup G(I)} [p/w \leq \tilde{e}_k] \ge 1- kq - \tau \text{ and } \Pr_{(p,w) \sim S(I) \cup G(I)} [p/w \geq \tilde{e}_k] \geq kq - \tau.
     \]
     Note that the above inequality also indicates that:
     \[
     \Pr_{(p,w) \sim S(I) \cup G(I)} [p/w > \tilde{e}_k] < kq + \tau \text{ and } \Pr_{(p,w) \sim S(I) \cup G(I)} [p/w < \tilde{e}_k] < 1- kq + \tau.
     \]
     Combining these inequalities gives us
     \begin{align*}
         &\Pr_{(p,w) \sim S(I) \cup G(I)} [p/w \geq \tilde{e}_{1}] \in [q - \tau, q + \tau]\\
         &\Pr_{(p,w) \sim S(I)\cup G(I)} [\tilde{e}_k > p/w \geq \tilde{e}_{k+1}] \in [q - 2\tau ,q + 2\tau] \tag{$1 \leq k \leq t-1$}\\
         &\Pr_{(p,w) \sim S(I)\cup G(I)} [\tilde{e}_{t} > p/w] \in [0, q + \tau).
     \end{align*}
The probability of drawing an item from $S(I) \cup G(I)$ is $p(S(I) \cup G(I))$. Recall that $L(I), S(I)$ and $G(I)$ are disjoint sets and their total profit are 1. Therefore, the probability of drawing an item from $S(I) \cup G(I)$ is $1 - p(L(I))$.  Now we condition on drawing an item from $I$ instead of $S(I) \cup G(I)$. Let $c = 1 - p(L(I))$. We have: 
     \begin{align*}
         &\Pr_{(p,w) \sim I} [p/w \geq \tilde{e}_{1} \text{ and } (p,w) \in S(I)\cup G(I)] \in [c (q - \tau), c (q + \tau)]\\
         &\Pr_{(p,w) \sim I} [\tilde{e}_k > p/w \geq \tilde{e}_{k+1} \text{ and } (p,w) \in S(I)\cup G(I)] \in [c (q - 2\tau) ,c(q + 2\tau)] \qquad k = 1, 2, \dots, t - 1\\
         &\Pr_{(p,w) \sim I} [\tilde{e}_{t} > p/w \text{ and } (p,w) \in S(I)\cup G(I)] \in [0, c(q + \tau)) 
     \end{align*}
Since the total profit of items in $S$ is normalized to 1 and items are sampled with probability equal to their profit, we can rewrite these inequalities as:
     \begin{align*}
         &p(\{(p,w)\in S(I) \cup G(I) \mid p/ w\geq \tilde{e}_1\})\in [c (q - \tau), c (q + \tau)]\\
        &p(\{(p,w)\in S(I) \cup G(I) \mid \tilde{e}_k >p/w\geq \tilde{e}_{k+1}\}) \in [c (q - 2\tau) ,c(q + 2\tau)] \tag{$1 \leq k \leq t-1$}\\
         &p(\{(p,w)\in S(I) \cup G(I) \mid \tilde{e}_t > p/ w\}) \in [0, c(q + \tau)),
     \end{align*}
     Substituting $q := \frac{\eps + \eps^{2}/2}{1 - p(L(\tilde{I}))}, \tau = \eps^{2}/5, c = 1 - p(L(I))$,  (note that $c\tau \leq \tau$), we get:
     \begin{align*}
         &p(\{(p,w)\in S(I) \cup G(I) \mid p/ w\geq \tilde{e}_1\})\in [\eps + 3\eps^{2}/10, \eps + 7\eps^{2}/10] \\
        &p(\{(p,w)\in S(I) \cup G(I) \mid \tilde{e}_k >p/w\geq \tilde{e}_{k+1}\})\ \in [\eps + \eps^{2}/10, \eps + 9\eps^{2}/10] \tag{$1 \leq k \leq t-1$}\\
         &p(\{(p,w)\in S(I) \cup G(I)\mid \tilde{e}_t > p/ w\}) \in [0, \eps + 7\eps^{2}/10)
     \end{align*}
     The efficiency of any item in $G(I)$ is at most $\eps^2$ and the total weight of items in $G(I)$ is at most 1. Therefore $p(G(I)) \leq \eps^2$. By the inequality above, we have:
 \begin{align*}
        &p(\{(p,w)\in S(I) \cup G(I) \mid \tilde{e}_{t-1} >p/w\geq \tilde{e}_{t}\}) \geq \eps + \eps^{2}/10. 
     \end{align*}
    Since $p(G(I)) \leq \epsilon^2 < \epsilon$, there must be an item from $S(I)$ in $\{(p,w)\in S(I) \cup G(I) \mid \tilde{e}_{t-1} >p/w\geq \tilde{e}_{t}\}$. This means $\tilde{e}_{t-1} \geq \eps^2$ as every item in $S(I)$ has efficiency at least $\epsilon^2$. 
    We can thus rewrite the previous set of inequalities as follows:
     \begin{align*}
         &p(\{(p,w)\in S(I) \mid p/ w\geq \tilde{e}_1\})\in [\eps + 3\eps^{2}/10, \eps + 7\eps^{2}/10] \subseteq [\eps, \eps + \eps^{2}) \\
        &p(\{(p,w)\in S(I) \mid \tilde{e}_k >p/w\geq \tilde{e}_{k+1}\})\ \in [\eps + \eps^{2}/10, \eps + 9\eps^{2}/10] \subseteq [\eps, \eps + \eps^{2}) \tag{$1 \leq k \leq t-2$}\\
        &p(\{(p,w)\in S(I) \cup G(I) \mid \tilde{e}_{t-1} >p/w\geq \tilde{e}_{t}\})\ \in [\eps + \eps^{2}/10, \eps + 9\eps^{2}/10] \\
         &p(\{(p,w)\in S(I) \cup G(I)\mid \tilde{e}_t > p/ w\}) \in [0, \eps + 7\eps^{2}/10)
     \end{align*}
     Given the condition on \cref{line:checkt}, there are two cases to consider:
     \begin{enumerate}
         \item $\tilde{e}_t < \eps^2, t' = t -1 $. By definition, no element in $S(I)$ has efficiency smaller than $\eps^2$ so we have:
     \begin{align*}
         \{(p,w)\in S(I) \mid \tilde{e}_{t-1} >p/w \}
         &= \{(p,w)\in S(I) \mid \tilde{e}_{t-1} >p/w \geq \tilde{e}_t\}\\
         &\subseteq \{(p,w)\in S(I) \cup G(I) \mid \tilde{e}_{t-1} >p/w\geq \tilde{e}_{t}\}\,.
    \end{align*}
    So 
    \[
         p(\{(p,w)\in S(I) \mid  \tilde{e}_{t-1} >p/w   \}) \in [0, \eps + 9\eps^2/10]\\
         \subseteq [0, \eps + \eps^2)
     \]
     \item $\tilde{e}_t \geq \eps^2, t' = t$. Then we have:
     \begin{align*}
                 p(\{(p,w)\in S(I) \cup G(I)\mid \tilde{e}_t > p/ w\}) \in [0, \eps + 7\eps^{2}/10)
                 \subseteq [0, \eps + \eps^{2})\,.
     \end{align*}
     \end{enumerate}
     Combining both cases, we get:
               \begin{align*}
         &p(\{(p,w)\in S(I) \mid p/ w\geq \tilde{e}_1\})\in [\eps , \eps + \eps^{2})\\
        &p(\{(p,w)\in S(I) \mid \tilde{e}_k >p/w\geq \tilde{e}_{k+1}\}) \in [\eps, \eps + \eps^{2})\tag{$1 \leq k \leq t'-1$}\\
         &p(\{(p,w)\in S(I) \mid \tilde{e}_{t'} > p/ w\}) \in [0, \eps + \eps^{2})\,.
     \end{align*}
     Conditioned on the sample $\vec{R}$ containing all elements of $L(I)$  after \cref{l:get:R}, $\tilde{e}_1, \tilde{e}_2, \dots, \tilde{e}_{t'}$ is an EPS with respect to $I$ as long as the following two conditions are satisfied:
     \begin{enumerate}
        \item $|\vec E|$ is at least the sample complexity of $\rQuantile$, 
        \item the approximate quantiles returned by $\rQuantile$ are all within error $\tau$.
    \end{enumerate}
    Now we consider the failure probability of condition (1).  Let $X$ be the random variable denoting the number of items in $\vec Q$ that are from $S(I) \cup G(I)$. Since the total profit of $S(I) \cup G(I)$ is $1 - p(L(I))$, the probability of getting an item from $S(I) \cup G(I)$ with weighted sampling is $(1 - p(L(I)))$, the expected number of items in $\vec Q$ that are from $S(I) \cup G(I)$ is $\mathbb{E}[X] = (1-p(L(I))) \cdot a \geq 3n_{rq}/2$. We consider the probability that $X \leq n_{rq}$. By Chernoff bound, we have
    \begin{align*}
        \Pr[X \leq n_{rq}] &= \Pr[X \leq (1-1/3)\mathbb{E}[X]]\leq e^{-\frac{\mathbb{E}[X]}{18}}\\
        &\leq e^{-\frac{n_{rq}}{12}}
        \leq e^{-\Omega(1/\rho^2)}
        = e^{-\Omega(1/\eps^4)}
        < \eps/3\,,
    \end{align*}
    where for the third inequality we used the sample complexity of $\rQuantile$ established in~\cref{theo:rquantile}. 
    $\vec E$ is obtained by removing all large items from $\vec Q$, and so $|\vec E| = X$. Therefore, $|\vec E|$ satisfies the sample complexity of $\rQuantile$ except with probability $\eps/3$.
    
     Conditioned on the success of (1), $\rQuantile$ runs at most $\eps^{-1}$ times and each run succeeds except with probability $\beta = \eps^2/36$, the probability that the approximate quantiles returned by $\rQuantile$ are all within error $\tau$ is at least $1 - \beta \eps^{-1} = 1 - \eps/36$. 
     
     By union bound, conditioned on the sample $\vec{R}$ containing all elements of $L(I)$  after \cref{l:get:R}, the probability that the approximate quantile sequence $\tilde{e}_1, \dots, \tilde{e}_{t'}$ in $\lcakp$ (\cref{alg:KP_LCA}) is an EPS with respect to $I$ is at least $1 - \eps/3 - \eps/36 = 1 - 13\eps/36$.
\end{proof}
\begin{lemma}\label{lem:mapping:feasible}
    The output $C$ of $\mappinggreedy$ (\cref{alg:convertgreedy}) is a feasible solution of $I$.
\end{lemma}
\begin{proof}
    We argue the feasibility of $C$, that is $w(C) \leq K$. First we consider the special case where $C = \{(p_{j+1}, w_{j+1})\}$. In this case $p_{j+1} > \sum_{i=1}^{j}p_i$, since all items in $S(\tilde{I})$ have the same profit $\eps^2$, $(p_{j+1}, w_{j+1})$ must be an item with profits greater than $\eps^{2}$, which implies $(p_{j+1}, w_{j+1}) \in L(\tilde{I})$. Since $L(\tilde{I}) \subseteq L(I)$ and all items in $I$ have weight at most $K$, $w(C) = w_{j+1} \leq K$ and so $C$ is feasible.
    
    Now we consider the general case where 
    \[C = \{(p,w) \in L(I) \mid p/w \ge \ p_{j}/w_{j}\} \cup \{(p,w) \in S(I) \mid p/w \ge \tilde{e}_{k-2} \}.\] Let $A_{0}(I), A_{1}(I), \dots, A_{t}(I)$ be the partition of items in $S(I)$ with respect to efficiency sequence $\tilde{e}_1, \tilde{e}_2, \dots,\tilde{e}_t$. Let $A_{0}(\tilde{I}), A_{1}(\tilde{I}), \dots, A_{t-1}(\tilde{I})$ be the partition of items in $S(\tilde{I})$ with respect to efficiency sequence $\tilde{e}_1, \tilde{e}_2, \dots,\tilde{e}_t$. By construction, $A_{i}(\tilde{I})$ contains $\lfloor \eps^{-1} \rfloor$ copies of item $(\eps^{2}, \eps^{2}/\tilde{e}_{i+1})$ for $i = 0, 1, \dots, t-1$. We consider the total weight of the items in $S(\tilde{I})$ that are included in the greedy solution $\greedyKP (\tilde{I})$. Since $\tilde{e}_{k} > p_j / w_j$, all items in $A_{i}(\tilde{I})$ must be included in the greedy solution $\greedyKP (\tilde{I})$, for $i = 0, 1, \dots, k-1$. Therefore,
    \begin{align*}
        w(S(\tilde{I}) \cap \greedyKP (\tilde{I})) &\geq \sum_{i = 0}^{k-1}w(A_{i}(\tilde{I}))\\
            &= \lfloor \eps^{-1} \rfloor \cdot \eps^{2}/\tilde{e}_{1} + \lfloor \eps^{-1} \rfloor \cdot \eps^{2}/\tilde{e}_{2} + \dots + \lfloor \eps^{-1} \rfloor \cdot \eps^{2}/\tilde{e}_{k}\\
            &= \lfloor \eps^{-1} \rfloor \eps^{2} \sum_{i = 1}^{k} (1/\tilde{e}_{i})\\
            &\geq (\eps^{-1} - 1) \eps^{2} \sum_{i = 1}^{k} (1/\tilde{e}_{i})\\
            &= (\eps + \eps^{2})\sum_{i = 1}^{k-2}(1/\tilde{e}_{i}) - 2\eps^{2}\sum_{i = 1}^{k-2}(1/\tilde{e}_{i}) + (\eps - \eps^{2})(1/\tilde{e}_{k-1} + 1/\tilde{e}_{k})\\
            &> (\eps + \eps^{2})\sum_{i = 1}^{k-2}(1/\tilde{e}_{i}) - 2\eps^{2}\sum_{i = 1}^{k-2}(1/\tilde{e}_{k-2}) + (\eps - \eps^{2})(1/\tilde{e}_{k-2} + 1/\tilde{e}_{k-2})\\
            &= (\eps + \eps^{2})\sum_{i = 1}^{k-2}(1/\tilde{e}_{i}) - 2\eps^{2}(k-1-\eps^{-1})/\tilde{e}_{k-2}.
    \end{align*}
    Since the total profit of the items in $I$ is 1 and $p(A_{i}(I)) \geq \eps$, the length of the efficiency sequence $t$ is at most $\lfloor \eps^{-1} \rfloor$. Since $k \leq t$ and $k < \lfloor \eps^{-1} \rfloor \leq \eps^{-1}$, combining both inequalities give us $(k - 1 - \eps^{-1}) < 0$. Therefore, 
    \[
     w(S(\tilde{I}) \cap \greedyKP (\tilde{I})) \geq (\eps + \eps^{2})\sum_{i = 1}^{k-2}(1/\tilde{e}_{i}) = \sum_{i = 1}^{k-2}((\eps + \eps^{2})/\tilde{e}_{i}) \geq \sum_{i = 0}^{k-3}w(A_{i}(I)).
    \]
    The last inequality is implied by the fact that $\tilde{e}_{i+1}$ lower bounds the efficiency of the items in $A_{i}(I)$ while the total profit of $A_{i}(I)$ is upper bounded by $\eps + \eps^2$. Now we consider the total weight of items in $L(\tilde{I})$ that are included in $\greedyKP (\tilde{I})$. Since $L(\tilde{I}) \subseteq L(I)$, we get $w(L(\tilde{I}) \cap \greedyKP (\tilde{I})) = w(L(I) \cap \greedyKP (\tilde{I}))$. Combining these inequalities, we have
    \begin{align*}
         w(C) &= w(\{(p,w) \in L(I) \mid p/w \ge \ p_{j}/w_{j}\}) +  w(\{(p,w) \in S(I) \mid p/w \ge \tilde{e}_{k-2} \}) \\
    &\leq w(L(\tilde{I}) \cap \greedyKP (\tilde{I})) + w((S(\tilde{I}) \cap \greedyKP (\tilde{I}))\\ 
    &= w(\greedyKP (\tilde{I})) \\
    &\leq \tilde{K}= K\,,
    \end{align*}
    concluding the proof.
\end{proof}
\begin{lemma}\label{lem:mapping_approx}
    If $\tilde{I}$ contains all large items from $I$, then $C$ is a $(1/2,6\eps)$-approximation solution of $I$.
\end{lemma}
\begin{proof}
    We consider the difference between the total profit of $C$ and $\greedyKP (\tilde{I})$. Let $\operatorname{\OPT}(I)$ be the optimal solution of instance $I$ and $\operatorname{\OPT}(\tilde{I})$ be the optimal solution of instance $\tilde{I}$. We first consider the case where $C = \{(p_{j+1}, w_{j+1})\}$. In this case $p(C) = p(\greedyKP (\tilde{I}))$. By~\cref{lem:relation:tilde:I},
    \begin{align*}
        p(C) &= p(\greedyKP (\tilde{I})) \geq 1/2p(\operatorname{\OPT}(\tilde{I})) \geq 1/2(p(\operatorname{\OPT}(I)) - 5\eps)\\
        &= 1/2p(\operatorname{\OPT}(I)) - 5\eps/2.
    \end{align*}
    
    Now we consider the case where $C = \{(p,w) \in L(I) \mid p/w \ge \ p_{j}/w_{j}\} \cup \{(p,w) \in S(I) \mid p/w \ge \tilde{e}_{k-2} \}$. Since we assume $\tilde{I}$ contains all \emph{large items} from $I$, therefore $\{(p,w) \in L(I) \mid p/w \ge \ p_{j}/w_{j}\} = \greedyKP (\tilde{I}) \cap L(\tilde{I})$. We consider the difference between the profit of $\greedyKP (\tilde{I}) \cap S(\tilde{I})$ and $\{(p,w) \in S(I) \mid p/w \ge \tilde{e}_{k-2} \}$. The least efficient item $(p_j, w_j)$ is either in $S(\tilde{I})$ or $L(\tilde{I})$. 
    
   \paragraph{Subcase 1:} If $(p_j, w_j) \in S(\tilde{I})$, then $\bigcup_{i=0}^{i=k-1}A_{i}(\tilde{I}) \subset \greedyKP (\tilde{I}) \cap S(\tilde{I}) \subseteq \bigcup_{i=0}^{i=k}A_{i}(\tilde{I})$. Since $p(A_{i}(I)) \in [\eps,\eps + \eps^{2})$ and $p(A_{i}(\tilde{I})) \leq \eps$ for $i = 0, 1, \dots, t-1$,
    \begin{align*}
        p(\{(p,w) \in S(I) \mid p/w \ge \tilde{e}_{k-2} \})&= \sum_{i=0}^{k-3}p(A_{i}(I)) \geq \sum_{i=0}^{k-3}p(A_{i}(\tilde{I}))\geq \sum_{i=0}^{k}p(A_{i}(\tilde{I})) - 3\eps\\
        &\geq p(\greedyKP (\tilde{I}) \cap S(\tilde{I})) - 3\eps.
    \end{align*}
    \paragraph{Subcase 2:} If $(p_j, w_j) \in L(\tilde{I})$, then $\bigcup_{i=0}^{i=k-1}A_{i}(\tilde{I}) = \greedyKP (\tilde{I}) \cap S(\tilde{I})$. Since $p(A_{i}(I)) \in [\eps,\eps + \eps^{2})$ and $p(A_{i}(\tilde{I})) \leq \eps$ for $i = 0, 1, \dots, t-1$,
    \begin{align*}
        p(\{(p,w) \in S(I) \mid p/w \ge \tilde{e}_{k-2} \})&= \sum_{i=0}^{k-3}p(A_{i}(I)) \geq \sum_{i=0}^{k-3}p(A_{i}(\tilde{I}))\geq \sum_{i=0}^{k-1}p(A_{i}(\tilde{I})) - 2\eps\\
        &= p(\greedyKP (\tilde{I}) \cap S(\tilde{I})) - 2\eps.
    \end{align*}
In both cases, $p(\{(p,w) \in S(I) \mid p/w \ge \tilde{e}_{k-2} \}) \geq p(\greedyKP (\tilde{I}) \cap S(\tilde{I})) - 3\eps$.
Using the assumption that $\tilde{I}$ contains $L(I)$, we obtain the following lower bound on $p(C)$:
\begin{align*}
    p(C) &= p(\{(p,w) \in L(I) \mid p/w \ge \ p_{j}/w_{j}\}) + p(\{(p,w) \in S(I) \mid p/w \ge \tilde{e}_{k-2} \})\\
    &\geq p(\greedyKP (\tilde{I}) \cap S(\tilde{I})) + p(\greedyKP (\tilde{I}) \cap L(\tilde{I}))- 3\eps\\
    &= p(\greedyKP (\tilde{I}))- 3\eps\\
    &\geq 1/2(p(\operatorname{\OPT}(\tilde{I}))) - 3\eps\\
    &\geq 1/2(p(\operatorname{\OPT}(I)) - 5\eps) - 3\eps\\
    &\geq 1/2p(\operatorname{\OPT}(I)) - 6\eps.
\end{align*}
Therefore, $C$ is a $(1/2,6\eps)$-approximation solution of $I$.
\end{proof}
\begin{lemma}[Consistency]\label{lem:lcakpconsistency}
$\lcakp$ provides query access consistent to a $(1/2,6\eps)$-approximation solution of $I$ with probability $1-\eps$. 
\end{lemma}
\begin{proof}
    The output of $\lcakp$ is determined by $\tilde{I}$ only. Thus, as long as two runs of $\lcakp$ constructs the same $\tilde{I}$, their answer for any index $i$ will be the same. Every run satisfying the following conditions constructs the same instance $\tilde{I}$:
    \begin{enumerate}
        \item all items in $L(I)$ are included in $\vec R$,
        \item $|\vec E|$ is at least the sample complexity of $\rQuantile$, 
        \item the approximate quantiles returned by $\rQuantile$ are both reproducible and within error $\tau$.
    \end{enumerate}
    We consider the failure probability of Condition (1). By~\cref{lem:large:item:weighted:sampling}, all items in $L(I)$ can be gathered by taking $m = O(\frac{1}{\eps^4}\log\frac{1}{\eps})$ samples with probability $1-\eps/3$. Therefore, the failure probability of Condition (1) is at most $\eps/3$. Conditioned on success of (1), he failure probability of Condition (2) is at most $\eps/3$ as proved in \cref{lem:EPSfailureprobability}.

    Conditioned on the success of (2), we consider Condition (3). Fix $k$ and internal randomness $r$, let $\vec{E_{1}} , \vec{ E_{2}}$ be two independent samples given to 
    $\rQuantile_{\rho, d, \tau, \beta}$ on two distinct runs of $\lcakp$. By the definition of reproducibility in \cite{impagliazzo2022reproducibility}, 
    \[\Pr_{\vec{E_1}, \vec{E_2}, r}[\rQuantile_{\rho, d, \tau, \beta}(\vec{E_1}, k) = \rQuantile_{\rho, d, \tau, \beta}(\vec{E_2}, k)] \geq 1 - \rho.
    \]
    Let $D_k$ be the set of all possible output of $\rQuantile_{\rho, d, \tau, \beta}(\vec E, k)$. Note that $D_k$ is finite since it shares the same domain with $\rQuantile_{\rho, d, \tau, \beta}(\vec E, k)$. Since $\vec{E_1}$ and $\vec{E_2}$ are two independent samples draw from $S(I) \cup G(I)$, we have:
    \begin{align*}
        &\Pr_{\vec{E_1}, \vec{E_2}, r}[\rQuantile_{\rho, d, \tau, \beta}(\vec{E_1}, k) = \rQuantile_{\rho, d, \tau, \beta}(\vec{E_2}, k)] \\
        = &\sum_{x \in D_k}\left(\Pr_{\vec E \sim S(I) \cup G(I)}[\rQuantile_{\rho, d, \tau, \beta}(\vec E, k) = x]\right)^{2}.
    \end{align*}
  Let $e_k \in D_k$ such that for any $x \in D_k$,
  \[
  \Pr_{\vec E \sim S(I) \cup G(I)}[\rQuantile_{\rho, d, \tau, \beta}(\vec E, k) = e_k] \geq \Pr_{\vec E \sim S(I) \cup G(I)}[\rQuantile_{\rho, d, \tau, \beta}(\vec E, k) = x].
  \]
  Let $y = \Pr_{\vec E \sim S(I) \cup G(I)}[\rQuantile_{\rho, d, \tau, \beta}(\vec E, k) = e_k] $, since
  \[
  \sum_{x \in D_k}\left(\Pr_{\vec E \sim S(I) \cup G(I)}[\rQuantile_{\rho, d, \tau, \beta}(\vec E, k) = x]\right) = 1.
  \]
Therefore,
  \[
  y^2 \cdot 1/y \geq \sum_{x \in D_k}\left(\Pr_{\vec E \sim S(I) \cup G(I)}[\rQuantile_{\rho, d, \tau, \beta}(\vec E, k) = x]\right)^{2} = 1 - \rho,
  \]
  and so
    \[\Pr_{\vec E, r}[\rQuantile_{\rho, d, \tau, \beta}(\vec E, k) = e_k] \geq 1 - \rho.
    \]
    The probability that $e_k$ is not a $\tau$-approximation quantile is at most $\beta$. By union bound $\rQuantile_{\rho, d, \tau, \beta}$ outputs some value $e_k$ and a $\tau$-approximation $(1-kq)$-quantile, except with probability $\rho + \beta$. Moreover, $\rQuantile_{\rho, d, \tau, \beta}$ is called $t$ times where $t= \lfloor q^{-1} \rfloor \leq q^{-1} = \frac{1-p(L(I))}{\eps + \eps^{2}/2} \leq \eps^{-1}$. By union bound, all calls to $\rQuantile$ are consistent to some $\tau$-approximation quantile sequence $e_1, e_2, \dots, e_{t'}$ except with probability $\eps^{-1}(\rho + \beta) = \eps/12$.

    By a union bound, all three conditions succeed except with probability $\eps/3 + \eps/3 + \eps/12 < \eps$.       
    By~\cref{lem:mapping_approx}, $\lcakp$ gives query access consistent to a $(1/2, 6\eps)$-approximation solution of $I$ with probability $1-\eps$.
\end{proof}
\begin{lemma}[Sample Complexity]\label{lem:lcakpsample}
    The sample complexity of $\lcakp$ is 
      \[
\left( 1/\eps \right)^{O(\log^{*}n)}.
    \]
\end{lemma}
\begin{proof}
    $\lcakp$ requires two samples to compute: $\vec Q$ and $\vec R$. On~\cref{line:checkifepspossible}, the algorithm checks if $1 - p(L(\tilde{I})) \geq \eps$, so  $|\vec Q| \leq \lceil \frac{3n_{rq}}{2\eps} \rceil$. As a result,
    \begin{align*}
        |\vec R| + |\vec Q|
        &= O\left(\frac{1}{\eps^4}\log \frac{1}{\eps}\right) + \frac{3}{2\eps} \cdot  \tilde{O}\left( \left( \frac{1}{\tau^2(\rho - \beta)^2}\right) \cdot \left(\frac{12}{\tau^2}\right)^{\log^{*}|\mathcal{X}| + 1} \right)\\
        &=\tilde{O}\left( \eps^{-7} \cdot \left( \eps^{-6} \right)^{O(\log^{*}n)}\right)\\
        &=\left( 1/\eps \right)^{O(\log^{*}n)},
    \end{align*}
    as desired.
\end{proof}
\noindent Combining~\cref{lem:lcakpconsistency} and~\cref{lem:lcakpsample} establishes~\cref{theo:lcakpresult}.

\section{Discussion and Future Work}

The obvious question left open by our work is whether the $\exp(O(\log^\ast n))$ dependence on the instance size can be improved, or removed altogether. Doing so would require an entirely different approach to deal with the consistency issue, as the reproducible median algorithm does require a dependence (albeit very mild) on the domain size.

The connection we made between LCAs and to reproducible algorithms was key to our main algorithmic result. We believe that exploring further this interplay, and how insights and techniques from reproducible learning algorithms can be used to design new or better LCAs for a variety of tasks, would be an fruitful direction of study.

Finally, as mentioned in the introduction, it would be interesting to see if the relaxation to \emph{average-case} local computation, as introduced in~\cite{BiswasCPR24}, would lead to either faster LCAs for Knapsack with weighted sampling access, or help circumvent our impossibility results absent this type of access.

\printbibliography

\end{document}